\documentclass[lettersize,conference]{IEEEtran}
\usepackage{amsmath,amsfonts}
\usepackage{algorithmic}
\usepackage{array}
\usepackage[caption=false,font=normalsize,labelfont=sf,textfont=sf]{subfig}
\usepackage{textcomp}
\usepackage{stfloats}
\usepackage{url}
\usepackage{verbatim}
\usepackage{graphicx}
\graphicspath{{./figs/}}
\def\BibTeX{{\rm B\kern-.05em{\sc i\kern-.025em b}\kern-.08em
    T\kern-.1667em\lower.7ex\hbox{E}\kern-.125emX}}
\usepackage{balance}
\usepackage{xcolor}
\usepackage[hidelinks]{hyperref}
\usepackage{notes2bib}
\usepackage{mathrsfs}

\usepackage[switch]{lineno}
\begin{document}

\title{Development Status of the KIPM Detector Consortium}

\author{
Dylan J Temples, %
Zo\"{e} J. Smith, %
Selby Q Dang, %
Taylor Aralis, %
Chi Cap, %
Clarence Chang, %
Yen-Yung Chang, \\%
%Selby Q Dang, \\
Maurice Garcia-Sciveres, %
Sunil Golwala, %
William Ho, %
Noah Kurinsky, %
Kungang Li, %
Xinran Li, %\\
Marharyta Lisovenko, \\%
Elizabeth Panner, %
Karthik Ramanathan, %
Shilin Ray, %
Brandon Sandoval, %\\%
Aritoki Suzuki, %
Gensheng Wang, \\%
Osmond Wen, %
Michael Williams, %
Junwen Robin Xiong, %
Volodymyr Yefremenko (KIPM Detector Consortium)%
\thanks{Please see the Acknowledgements section of this article for author affiliations.}
% \thanks{D.J. Temples is with the Emerging Technologies Directorate, Fermi National Accelerator Laboratory, Batavia, IL, USA.} %
% \thanks{W. Ho and K. Ramanathan are with the Department of Physics, Washington University in St. Louis, St. Louis, MO, USA.} %
% \thanks{T. Aralis, S.Q. Dang, N. Kurinsky, Z. Smith, and O. Wen, are with the Kavli Institute for Particle Astrophysics and Cosmology, SLAC National Laboratory, Menlo Park, CA, USA. S. Q. Dang and Z. Smith are also with the Department of Physics, Stanford University, Stanford, CA, USA.} %
% \thanks{Y.-Y. Chang, M. Garcia-Sciveres, K. Li, X. Li, A. Suzuki, and M. Williams are with the Division of Physics, Lawrence Berkeley National Laboratory, Berkeley, CA, USA.} %
% \thanks{S. Golwala, C. Cap, S. Ray, B. Sandoval, and J.R. Xiong are with the Division of Physics, Mathematics, and Astronomy, California Institute of Technology, Pasadena, CA, USA.} %
% \thanks{G. Wang, M. Lisovenko, and V. Yefremenko are with %the Division of Physical Sciences \& Engineering, 
% Argonne National Laboratory, Lemont, IL, USA.} % 
% \thanks{C. Chang is with the Kavli Institute for Cosmological Physics, University of Chicago, Chicago, IL, USA.} %
% %
% %
% \thanks{E. Panner is with the Department of Physics and Astronomy, Tufts University, Medford, MA, USA.}
%
%   
}

\markboth{IEEE Transactions on Applied Superconductivity,~Vol.~X, No.~Y, September~2025}%
{Dylan J Temples, \MakeLowercase{\textit{(et al.)}: The KIPM Detector Consortium}}

\maketitle

\begin{abstract}
A Kinetic Inductance Phonon-Mediated Detector is a calorimeter that uses kinetic inductance detectors (KIDs) to read out phonon signals from the device substrate. We have established a consortium comprising university and national lab groups dedicated to advancing the state of the art in these detectors, with the ultimate goal of designing a detector with sub-eV threshold on energy deposited in the substrate, enabling searches for both light dark matter and low-energy neutrino interactions. This consortium brings together experts in kinetic inductance detector design, phonon and quasiparticle dynamics, and noise modeling, along with specialized fabrication facilities, test platforms, and unique calibration capabilities. Recently, our consortium has demonstrated a resolution on energy absorbed by the sensor of 2.1 eV, the current record for phonon-sensitive KIDs, though due to $\sim$$1\%$ phonon collection efficiency, the overall detector resolution was limited to 320 eV. The current focus of the consortium is modeling and improving the phonon collection efficiency and implementing low-$\boldsymbol{T_c}$ superconductors, both of which serve to improve the overall energy resolution and threshold of the detectors. 

\noindent\textbf{FNAL Report Number: FERMILAB-PUB-25-0652-ETD}
\end{abstract}

\begin{IEEEkeywords}
Calorimetry, dark matter, phonons, superconducting detectors, superconducting resonators
\end{IEEEkeywords}

%%%% ==== ============ ==== %%%%
%%%% ==== INTRODUCTION ==== %%%%
%%%% ==== ============ ==== %%%%

\section{Introduction}
\begin{figure}[!t]
\centering
\includegraphics[width=\linewidth]{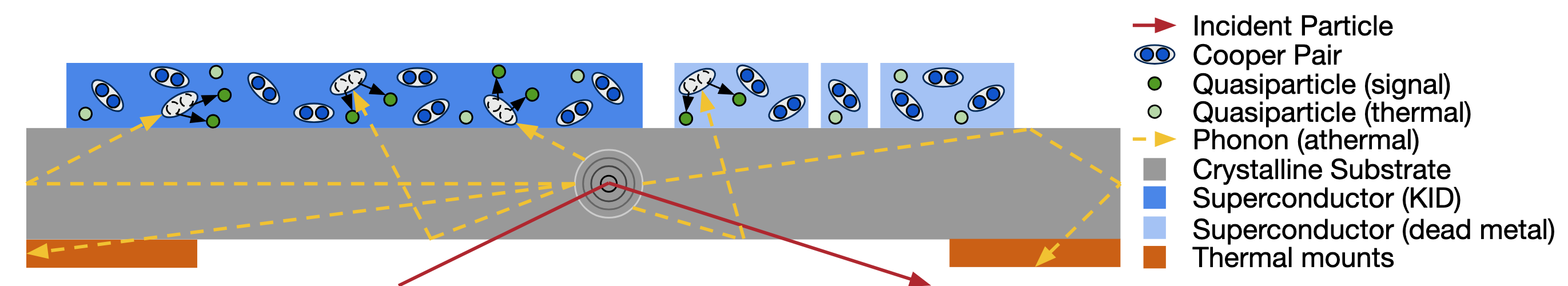}
\caption{Working principle for the KIPM detector, in which a particle scattering event produces athermal phonons that are either absorbed by the sensitive superconducting element (the inductor) or lost to other superconducting structures, device mounts, or to downconversion. }
\label{fig:devicecartoon}
\end{figure}

\IEEEPARstart{A}{} Kinetic Inductance Phonon-Mediated (KIPM) Detector is a calorimeter consisting of a crystalline target (Si to date, with Ge and other materials planned) and a number of superconducting resonators (kinetic inductance detectors, KIDs) patterned onto its surface. A particle scattering or absorption event in the device substrate produces a burst of athermal phonons with $\mathscr{O}$(meV) energies that propagate quasiballistically through the substrate~\cite{Nakamura2010, Irwin1995} until they are absorbed by the superconductor forming the sensitive element (the KID inductor) or are lost to insensitive (``dead") metal, mounting structures, or downconversion below the Cooper-Pair binding energy ($\approx 400~\mu\mathrm{eV}$ in thin-film Al). Absorbed phonons break Cooper pairs in the inductor, transiently changing its complex surface impedance and thus the KID's resonance frequency $f_r$ and quality factor $Q_i$ until the created quasiparticles (QPs) recombine into pairs. These changes are tracked by measuring the forward scattering parameter $S_{21}$ of a continuous-wave radiofrequency (RF) stimulus at the KID's quiescent $f_r$. Fig.~\ref{fig:devicecartoon} shows a schematic of these processes.

KIPM detectors offer an attractive architecture for phonon sensing, specifically in the field of dark matter (DM) direct detection (scattering and absorption), as they are natively suited for frequency-domain multiplexing. They are also non-dissipative, and as such are amenable to quantum information science techniques including squeezed amplification.
While currently outperformed in energy resolution by phonon-mediated detectors using transition-edge sensors (TESs)~\cite{Chang2025}, the above considerations, the inherently dissipative nature of TESs (which limits their performance), and the potentially simpler KID fabrication process, motivate our consortium to improve KIPM detectors to match or exceed the performance of TES-based devices. 

The structure of this article is as follows. In $\S$\ref{sec:current_status}, we summarize the results of testing our current device architecture. Our near-term plans for improving the performance of this design are presented in $\S$\ref{sec:near_term_improvements}. In $\S$\ref{sec:lowTc} and $\S$\ref{sec:paakipm} we motivate the use of low-$T_c$ superconductors and summarize a design of a novel phonon-absorber-assisted (PAA) KIPM detector, respectively. Finally, in $\S$\ref{sec:capabilities}, we review the fabrication and test facilities available within the consortium before concluding in $\S$\ref{sec:conclusion}.

%%%% ==== ============== ==== %%%%
%%%% ==== CURRENT STATUS ==== %%%%
%%%% ==== ============== ==== %%%%

\section{Current Status} \label{sec:current_status}
\begin{figure}[!t]
\centering
\includegraphics[width=1.0\linewidth]{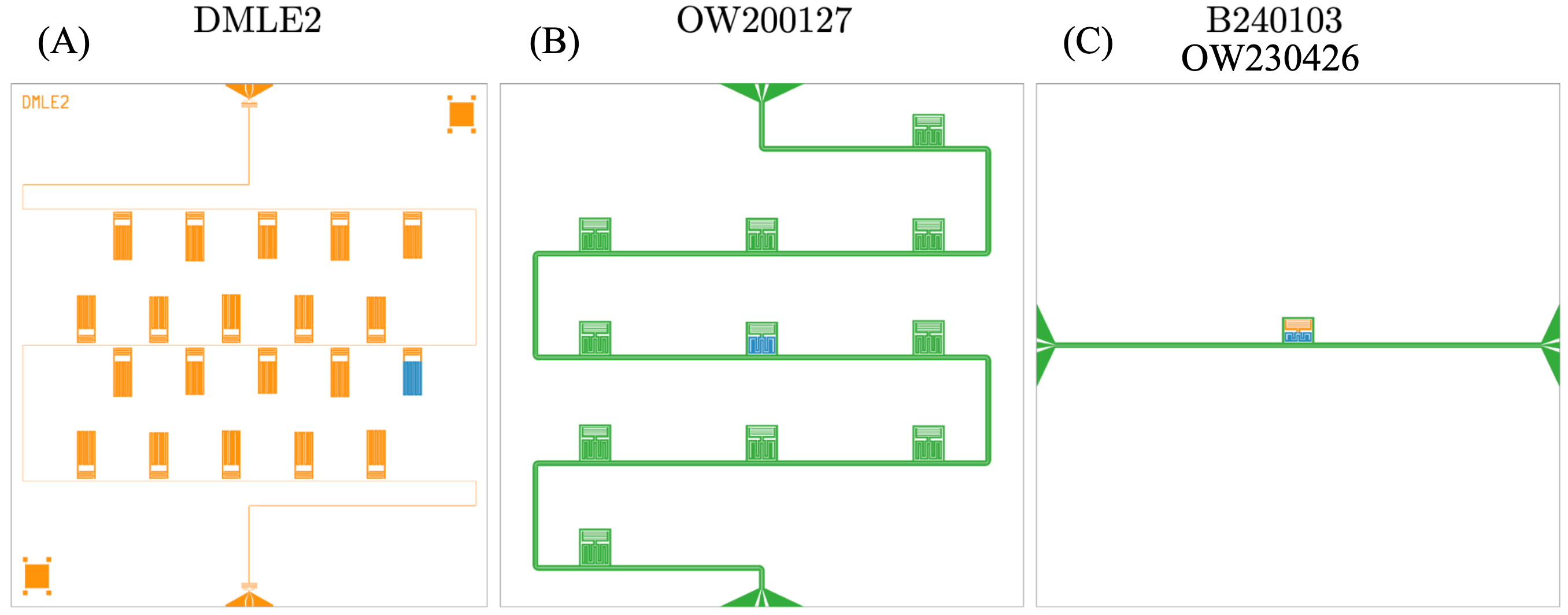}
\caption{Three KIPM detector design masks (labeled A, B, and C) and device names studied within our consortium, all of which feature a (2.2 cm)$^2$ silicon substrate. Orange is ``dead" (insensitive) Al, green is Nb, and blue shows the sensitive Al element for which $\eta$ is evaluated. Adapted from~\cite{Wen2025}.}
\label{fig:device-masks}
\end{figure}

\noindent Our consortium has recently demonstrated world-leading KIPM detector resolution on energy in the sensor (\textit{i.e.}, energy resolution of the quasiparticle channel) of $\sigma_{E_\mathrm{abs}}= 2.1$ eV~\cite{Temples2024, Cardani2018, Cardani2021a, Cruciani2022, Delicato2023}. However, our resolution on energy deposited in the substrate (\textit{i.e.}, energy resolution of the phonon channel) is limited to $\sigma_{E_\mathrm{dep}} \approx 320$ eV due to our observed single-percent phonon collection efficiency, $\eta$ (see \S\ref{sec:pulses} for a discussion on the measurement methodology). Similar values of $\eta$ have been observed in all of our current devices, as summarized in Table~\ref{tab:current_status} and discussed in \S\ref{sec:pce}. In this table, the first two columns indicates the mask ID and device name. The third column indicates the fractional area coverage of the sensitive inductor $f_\mathrm{fill}^\mathrm{ind}$ with respect to a single face of the chip~\bibnote{To determine the total fill fraction, divide the quoted number by 2. This neglects the 4 thin faces, but their contribution to the total surface area of the device is negligible.}. The following three columns show the device resolution on energy deposited in the substrate $\sigma_{E_\mathrm{dep}}$, the phonon collection efficiency $\eta$, and the resolution on energy absorbed by the inductor $\sigma_{E_\mathrm{abs}}$, respectively. The final two columns indicate which source of noise dominates the device and the relation between the internal quality factor $Q_i$ and the coupling quality factor $Q_c$. The first two rows are for the same device~\cite{Moore2012a}, when considering all resonators or a single resonator, for comparison to single-resonator devices. Another Mask C device is under test (OW230426), with the results of this calibration left to a future publication. The design masks of these devices are shown in Fig.~\ref{fig:device-masks}. 

\begin{table*}
\centering
\caption{Summary of test results from three devices of current resonator architecture}
\begin{tabular}{|clcccccl|}
\hline
Mask  & Device & $f_\mathrm{fill}^\mathrm{ind}$    & $\sigma_{E_\mathrm{dep}}$ {[}eV{]}     & $\eta$ {[}\%{]}         & $\sigma_{E_\mathrm{abs}}$ {[}eV{]}    & Dominant Noise & $Q$ Relation  \\
\hline
(A) & DMLE2~\cite{Moore2012a}     & $4.6\times10^{-2}$  & 380    & 7.0 $\pm$ 1.1    & 5.9 $\pm$ 0.9         & Amplifier      & $Q_i \gg Q_c \approx 10^4$                 \\
    & DMLE2 (1)~\cite{Wen2025}    & $2.3\times10^{-3}$  & 540    & 1.1 $\pm$ 0.3    & 5.9 $\pm$ 0.9         & Amplifier      & $Q_i \gg Q_c \approx 10^4$                 \\
(B) & OW200127~\cite{Temples2024} & $1.4\times10^{-3}$  & 320    & 0.78 $\pm$ 0.07  & 2.1 $\pm$ 0.1         & TLS            & $Q_i \approx 0.6Q_c \approx 4 \times 10^5$ \\
(C) & B240103~\cite{Wen2025}      & $8.2\times10^{-4}$  & 850    & 1.1 $\pm$ 0.2    & 23                    & Amplifier      & $Q_c \gg Q_i \approx 3 \times 10^4$    \\    
\hline
\end{tabular}
\vspace{0.2em}
\label{tab:current_status}
\end{table*}

From these results, we can qualitatively describe the successes and failures of our current KIPM detector architecture, which are largely consistent across all devices. Overall, we observe high resonator quality factors with $Q_i > 10^4$ (often $>10^5$), which, coupled with the millisecond-scale pulse lifetimes~\cite{Temples2024} enables the obtained sensor resolution. This resolution is, however, limited by the presence of noise due to electrically-active two-level systems (TLSs), with power spectral density $J_\mathrm{TLS}(1~\mathrm{kHz}) \approx 3\times 10^{-20}$ Hz$^{-1}$ for $\approx 3 \times 10^{6}$ readout photons stored in the resonator, a level observed in all resonators of our current design regardless of capacitor material choice~\cite{Wen2025}. This TLS noise power is comparable to what is observed in KIDs of various designs and materials~\cite{Zmuidzinas2012}. In multiple KIPM detectors of different materials and resonator design within our consortium, we have observed non-trivial phonon pulse shapes with interesting temperature dependence of both fall times and peak amplitudes, %The pulse shape is 
discussed in \S\ref{sec:pulses}.

These features drive the near- and far-term optimization of future designs for both the resonator and device layout, as well as the ongoing measurements to understand our energy collection and signal production.

\subsection{Phonon Collection Efficiency} \label{sec:pce}

The primary limiting factor in achieving eV-scale resolution with our devices is the single-percent $\eta$. To maximize $\eta$, we focus on understanding the current sources of phonon loss, which we categorize as follows: (1) losses to Cooper-Pair-breaking inefficiency; (2) a general, non-superconductor loss (including loss to normal metal mounting structures and surface-mediated downconversion); and (3) losses to ``dead,'' or inactive, superconducting structures deposited on the device substrate (including the RF feedline and the KID capacitor). 

The category (1) loss is $1-\eta_\mathrm{qp}$ where $\eta_\mathrm{qp}=0.46$~\cite{Guruswamy2014, Day2024, Wen2025}, for Al absorbers, is the efficiency for the quasiparticle to retain the absorbed energy, the remainder of which is lost to sub-gap phonons emitted as QPs relax to near the superconducting gap energy. Losses in categories (2) and (3) are device-dependent.  

\begin{figure}[!t]
\centering
\includegraphics[width=0.9\linewidth]{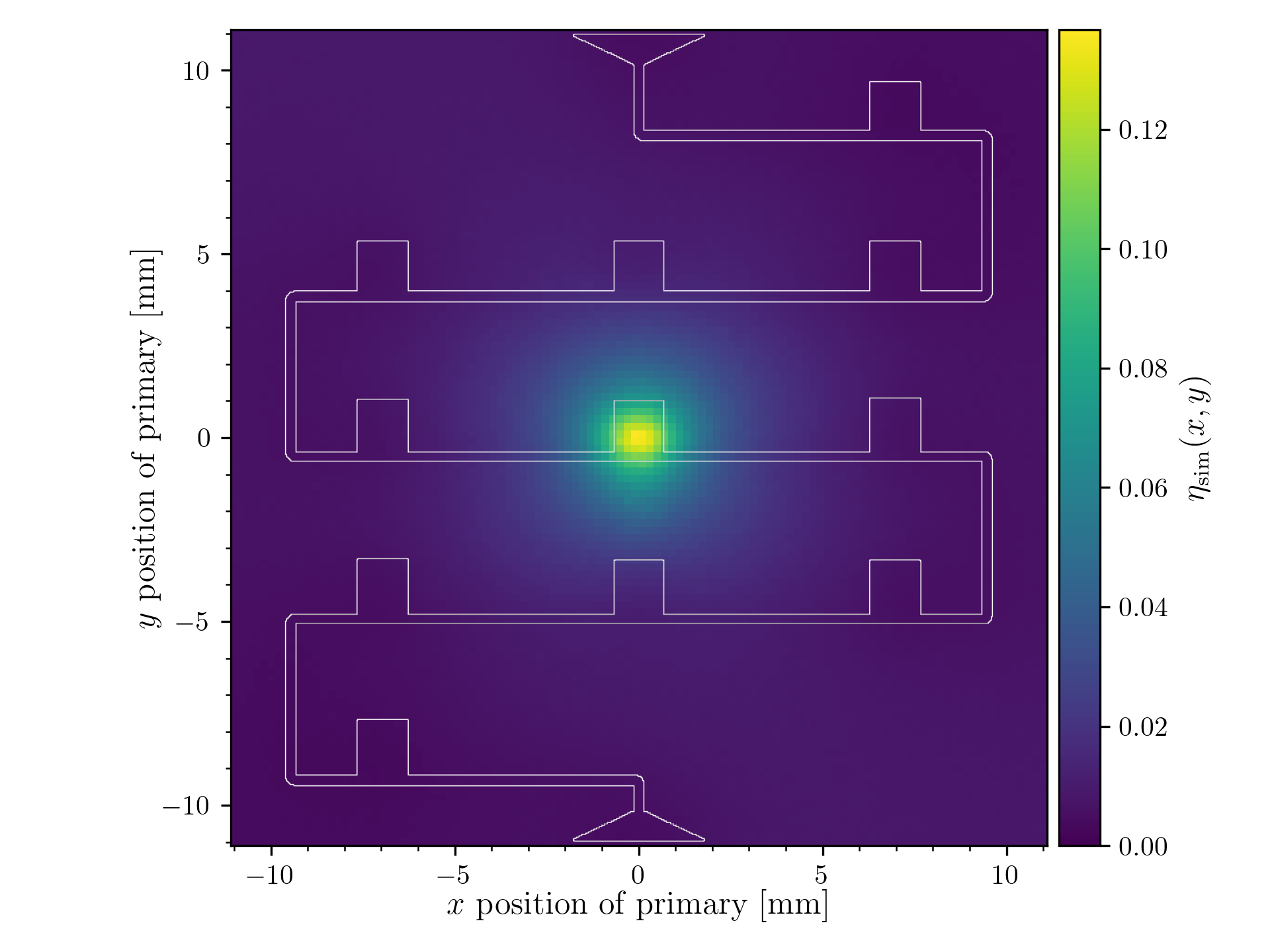} \\
\caption{The simulated phonon collection efficiency $\eta^\mathrm{sim}(x,y)$ as a function of position across a KIPM detector with design Mask B is shown by the color scale. The white outline indicates the location of the superconducting features.}
\label{fig:fnal-I-sims}
\end{figure}

We are able to investigate these loss mechanisms both in experiment and in simulation to understand their relative importance and optimize device design. In Fig.~\ref{fig:fnal-I-sims} (top), we show the simulated position-dependent phonon collection efficiency $\eta^\mathrm{sim}(x,y)$ for a Mask B device modeled using the Geant4 Condensed Matter Physics (G4CMP) framework ~\cite{Kelsey2023}. We generate $10^7$ electron-hole (e$^{-}$/h$^+$) pairs, each with initial energy (kinetic and band gap) corresponding to the photon wavelength used in experiment (470 nm), in the substrate with uniform density in $(x,y)$. The $z$ position~\bibnote{With $z=0$ being the face opposite that with the MKID structures.} of each pair is drawn from an exponential distribution parameterized by the attenuation length of 470-nm light in Si, 0.602 $\mu$m~\cite{GREEN20081305}. We use a simple model of phonon absorption and reflection at substrate-superconductor interfaces (with absorption probabilities of aluminum and niobium set to $p_\mathrm{abs}^\mathrm{Al}=0.488$ and $p_\mathrm{abs}^\mathrm{Nb}=0.745$, respectively~\cite{Yelton2024}), and disregard phonon loss to mounting surfaces. To determine collection efficiency $\eta^\mathrm{sim}(x,y)$, we tally the total phonon energy absorbed by the target inductor volume from events originating in a 2D bin centered at $(x,y)$ and divide by the total energy originating in that bin.

The calibration of a Mask B device (OW200127) ~\cite{Temples2024} was done with a fiber pointed at the rear of the substrate beneath the central Al resonator. As such, we have only a single point of comparison between simulation and experiment. We average $\eta^\mathrm{sim}(x,y)$ %the phonon collection efficiency 
over a circle with radius given by the expected laser spot size (0.5 mm) centered on the chip below the phonon-sensitive resonator, and find it to be $\bar{\eta}^\mathrm{sim}_{\mathrm{spot}}=(9.43\pm0.05)$\%. This value is much larger than the measured value of $\eta=0.78\%$. Some discrepancy is to be expected, as many known effects were not accounted for in the simulation:
\begin{enumerate}
    \item The pair-breaking efficiency $\eta_\mathrm{qp}$ is neglected. Accounting for this factor reduces the value of $\bar{\eta}^\mathrm{sim}_{\mathrm{spot}}$ by roughly half.
    \item Loss to normal-metal mounting structures has been disabled, which would serve to lower $\eta^\mathrm{sim}(x,y)$ primarily near the corners where the device is mounted.
    \item We find that if we displace the center of the spot by 1 mm, emulating a small pointing or alignment offset between the optical fiber used to deliver the photon pulses and the center of the inductor, the value of $\bar{\eta}^\mathrm{sim}_{\mathrm{spot}}$ is lowered by 61\%. Other data --- from larger-diameter devices with tens of resonators that measure the position dependence of the signal~\cite{Moore2012a}, and from a measurement in which the deposition position is scanned across the back of the device~\cite{ZoeMEMS2025} using a cryogenic, steerable optical calibration system~\cite{Stifter2024,Tabassum2025}, imply a much weaker dependence of $\eta$ on event position, suggesting $p^\mathrm{Al}_\mathrm{abs}$ (and possibly $p^\mathrm{Nb}_\mathrm{abs}$) is much lower than measured in~\cite{Yelton2024}. 
\end{enumerate}
There are other processes that are either not modeled, or not modeled sufficiently in this simulation that impact $\eta$ and may explain the lingering discrepancy between simulation and data:
\begin{enumerate}
    \item \textbf{Phonon recycling:} phonon reemission from recombining QPs. Primarily, this would serve to (a) increase the phonon collection efficiency, and (b) lengthen the timescale of phonon arrival.
    \item \textbf{Surface-mediated downconversion:} Both the polished and unpolished faces of the substrate may, upon reflection, induce phonon downconversion to energies below $2\Delta$, diminishing $\eta$.
    \item \textbf{Absorption probabilities:} While the values chosen for $p_\mathrm{abs}^\mathrm{Nb}$ and $p_\mathrm{abs}^\mathrm{Al}$ were motivated by Ref.~\cite{Yelton2024}, differences in film thickness and fabrication process may lead to substantially different values. 
\end{enumerate}
Until recently, effect (1) was not possible in the simulation framework but has been added in a recent release and can now be incorporated. This allows each effect to be investigated in both simulation and experiment, primarily by investigating the position dependence of phonon collection efficiency and pulse shapes using the cryogenic steerable optical calibration platform, then comparing that to simulations in which parameters are tuned to minimize discrepancy with data.

We undertake measurements of phonon collection efficiency and energy resolution (on energy deposited in the substrate and absorbed in the sensor) using a pulsed LED to deposit optical photons in the substrate~\cite{Cardani2018,Cruciani2022,Temples2024}. This method has many advantages over radioactive sources: it can be turned off to acquire uncontaminated noise data, the energy deposition can be varied over a wide range, from single photon to many keV, and at high energies, the variance on the energy measured is dominated by photon Poisson statistics, which give the number of photons absorbed and thus the absolute energy deposited. (Once near-single-photon energy resolution is achieved, the number of photons absorbed will be clear from the data.)

We use an optimal-filter (OF) based estimator for energy~\cite{Golwala2000}. When estimating absorbed energy, this approach currently makes the assumption that the peak pulse height, converted to QP density, gives the absorbed energy via $E_{abs} = n_{qp} V \Delta$.  Interpretation of pulse height is more complicated when the phonon energy arrives over a time scale comparable to or longer than the QP recombination time, which may occur if reabsorption of recombination phonons is significant.

\subsection{Pulse Shape and Lifetime} \label{sec:pulses}
The resonator response in the time domain to phonon signals yields pulses not described by a single exponential fall-time constant but rather a two-component model comprising a prompt component and a delayed component, each with an independent exponential rise- and fall-time, as introduced in~\cite{Moore2012a} and given physical interpretation in~\cite{Temples2024}, which proposes that at low temperatures the prompt(delayed) decay time represents the phonon(quasiparticle) lifetime and that these roles are reversed at high temperatures, with the crossover occurring at $\approx110$ mK. This pulse shape has been observed in multiple devices in different test facilities, and is believed to be a consequence of complex interplay between phonon and QP dynamics.

Armed with this model, we revisit the interpretation of the pulsed-LED/fiber energy calibration described above. When the spot is located immediately opposite the resonator on the bottom face, the prompt component dominates the pulse amplitude because of the relatively large solid angle subtended by the inductor with respect to this deposition location: that part of the pulse is due to phonons that have undergone only a few reflections before reaching the inductor. The energy resolution and $\eta$ inferred for this deposition location is a combination of the values for the first ``wave" of phonons and for the later, uniformly distributed component. The latter component still has a role because of the OF construction, combining information on all time scales. Conversely, a similar calibration with the fiber positioned far from the resonator involves sensing primarily of the delayed component because now the inductor subtends a small solid angle at the energy deposition location and the phonons must undergo many reflections to reach the resonator. The energy resolution and $\eta$ thus obtained are effectively only for the delayed component, which is sensitive to phonon loss to mounting structures or other superconducting films where the prompt component was not. Phonons absorbed by the latter (with potentially higher $T_c$, \textit{e.g.}, Nb) can later be reemitted into the substrate via QP recombination and further contribute to the timescale of the delayed component (phonon recycling).

\begin{figure}[!t]
\centering
\includegraphics[width=\linewidth]{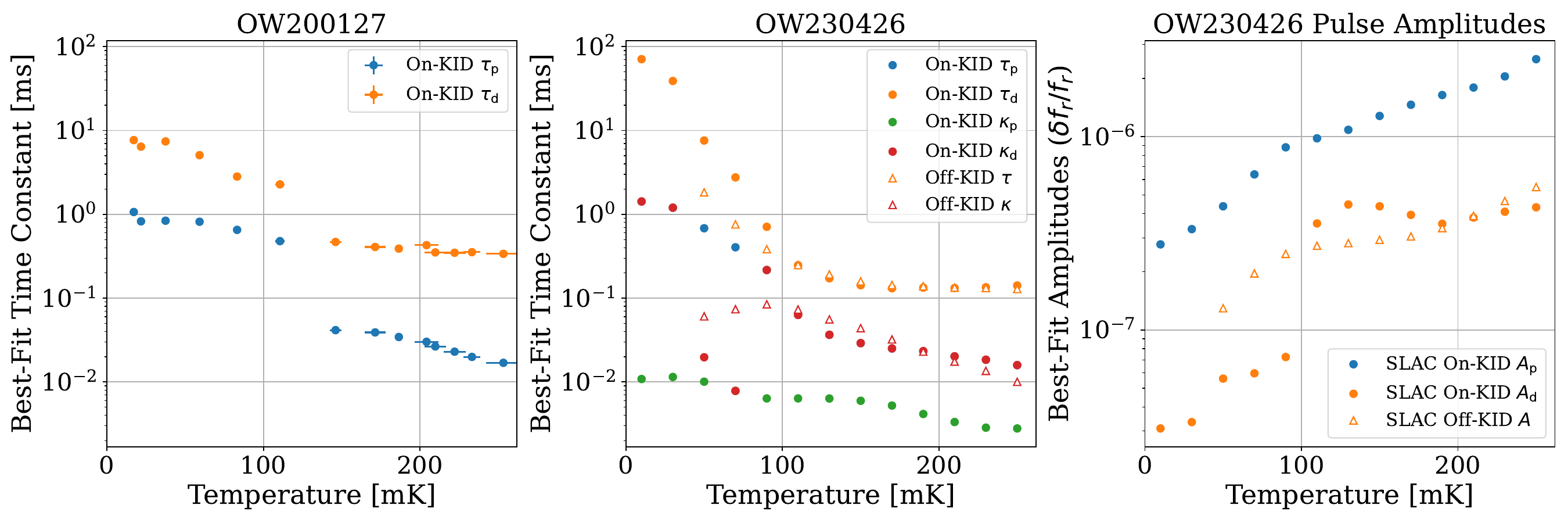}
\caption{Comparison of best-fit pulse model parameters as a function of device temperature between the OW200127 (Mask B, left) and OW230426 (Mask C, middle \& right) devices. The data for OW200127 (reproduced from~\cite{Temples2024}) are the best-fit fall-time constants for the two-component pulse shape model. The OW230426 plots contain both On-KID data, fit to the two-component model (solid circles), and Off-KID data which are well described by a pulse with a single exponential fall-time (open triangles), except for the two lowest temperatures which are omitted from this plot. Where not visible, the On-KID $\tau_\mathrm{p}$ points fall behind the On-KID $\kappa_\mathrm{d}$ points, highly suggestive of phonon recycling. (Right) Best-fit pulse amplitudes for the OW230426 device for the on-KID two-component model (solid circles) and off-KID one-component model (open triangles).}
\label{fig:pulse_temperature}
\end{figure}

To investigate this partitioning of phonon energy between prompt and delayed components, we compare pulsed LED measurements of a Mask C device (OW230426) at SLAC with the fiber in two positions: just beneath the resonator (``on-KID'') and a position displaced from the center of the chip by 8.5 mm vertically and 9.7 mm horizontally (``off-KID''). At each position, we collect $\approx25$ pulses in the $\delta f/f$ readout quadrature and average them. These average pulses are then fit to the two-component pulse shape model~\cite{Moore2012a,Temples2024} 
\begin{align}
    A(t) =A_p (e^{-t/\tau_\mathrm{p}}-e^{-t/\kappa_\mathrm{p}})  + A_d (e^{-t/\tau_\mathrm{d}}-e^{-t/\kappa_\mathrm{d}}) 
\end{align}
to extract the rise time $\kappa_{\mathrm{p(d)}}$, fall time $\tau_{\mathrm{p(d)}}$, and amplitude $A_\mathrm{p(d)}$ of the prompt(delayed) component. We also fit the off-KID data to a model with only a single amplitude ($A$), rise time ($\kappa$), and fall time ($\tau$), as this captures the pulse shape better than the two-component model. We investigate these parameters as a function of device temperature and compare them to measurements made with a Mask B device (OW200127) for a fiber in the ``on-KID" position~\cite{Temples2024}, the results of which are shown in Fig.~\ref{fig:pulse_temperature}.

We see qualitative agreement in the behavior of $\tau_\mathrm{p}$ and $\tau_\mathrm{d}$ versus temperature between on-KID calibrations of OW230426 and OW200127. The decay time of the prompt component ($\tau_\mathrm{p}$) is of the same scale in both devices, asymptoting to $\mathscr{O}(100)~\mu$s, though the OW200127 data show a sharper increase in $\tau_\mathrm{p}$ in the 120-130 mK regime but a flatter temperature dependence below that temperature. At temperatures above 100 mK, the off-KID best-fit time constants follow the time constants of the on-KID delayed component, as does the amplitude of the on-KID delayed component ($A_\mathrm{d}$) and the off-KID amplitude ($A$). This is a strong indicator that pulses for events originating far from the resonator are comprised primarily of the delayed phonon component. 

\begin{figure}[!t]
\centering
\includegraphics[width=\linewidth]{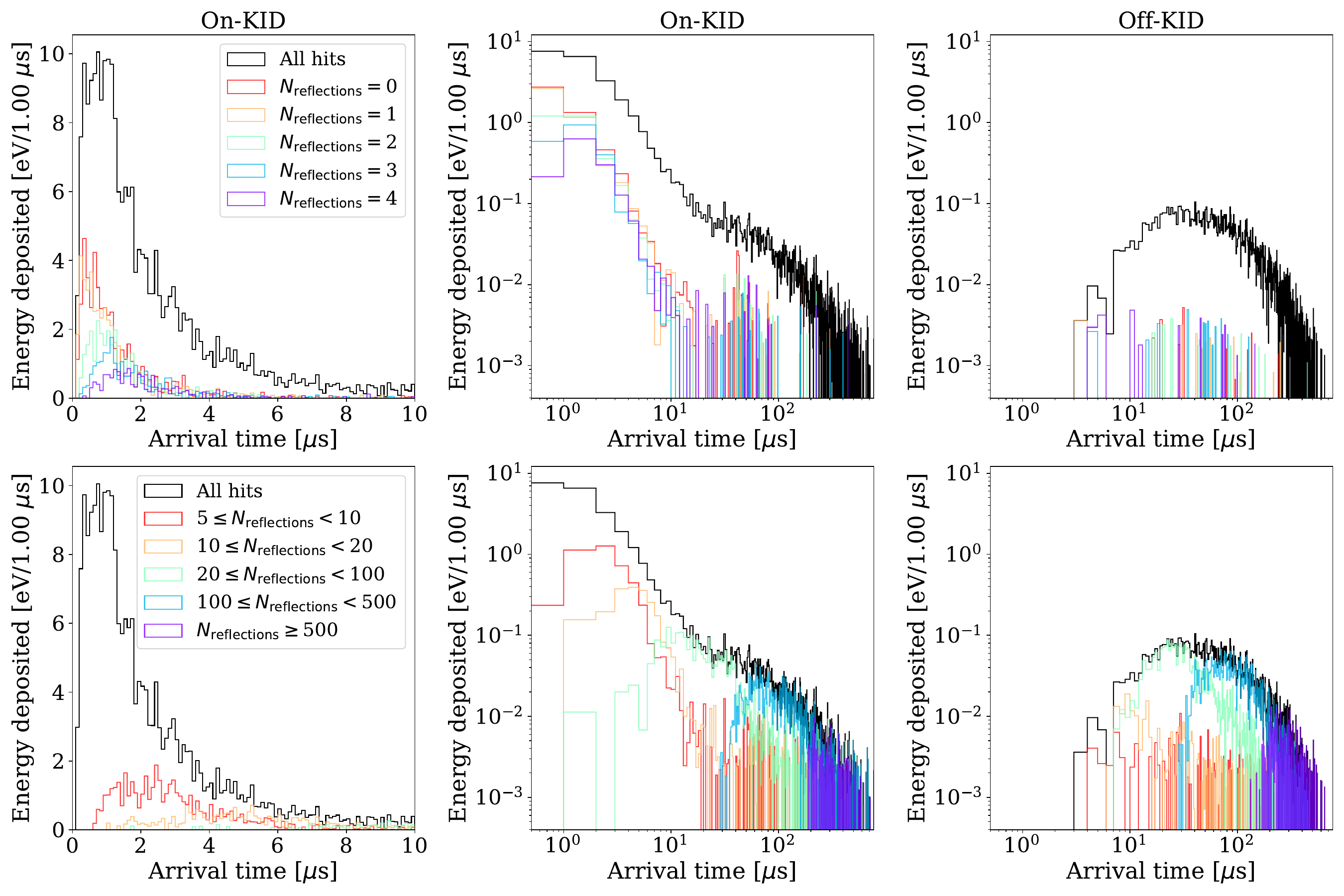} 
\caption{Histogram of simulated phonon energies binned by arrival times on the inductor for the Mask C design. The colors indicate the time distribution of phonons that have undergone a specified number of reflections before being absorbed by the inductor. The total absorbed energy (per $\mu$s) for phonons of undergoing any number of reflections is shown by the black line. The top row shows the distributions of arrival times for phonons that have reflected fewer than five times, while the bottom row is for phonons that have reflected five or more times. (Left column) ``On-KID" arrival times in the first 10 $\mu$s of the event. (Middle column) ``On-KID" arrival times up to 800 $\mu$s. (Right column) ``Off-KID" arrival times  up to 800 $\mu$s. The disparity in timescale for on- and off-KID distributions demonstrates the dominance of the delayed component for events far from the inductor, as well as the dominance of the prompt component for events originating near the inductor. Note that despite the different bin width of the left column, the units of the vertical axis have been scaled to match that of the other two so that energy deposition rates can be directly compared across all plots.}
\label{fig:phonon-times}
\end{figure}

To support the argument of the partitioning of phonon energy into prompt and delayed components based on the solid angle subtended by the inductor with respect to the position of the incident particle, we again turn to simulation, which allows us to track the history of individual phonons. In a simulation of Mask C, using the same configuration as the simulation of Mask B (\S\ref{sec:pce})~\bibnote{Though with the initial energy of the e$^-$/h$^+$ pair lowered to mimic the energy imparted by the 625 nm laser, and the absorption length parameter increased accordingly to 3 $\mu$m.}, we generate e$^-$/h$^+$ pairs in ``spots" at the same two locations as investigated in the above experiment. In Fig.~\ref{fig:phonon-times}, we histogram the energy absorbed by the resonator as a function of phonon arrival time, separated by the number of reflections the absorbed phonon underwent. In this, we see the prompt component ($\le$ 10~$\mu$s), probed by the ``on-KID" events, is primarily dominated by phonons that have reflected fewer than ten times. Conversely, the ``off-KID" distribution indicates there is very little prompt component, and the pulses are primarily generated by phonons that have undergone many ($>$20) reflections, having propagated away from the interaction site. Note that the delayed component arises in both On- and Off-KID events at similar timescales and amplitude, in agreement with the interpretation of Fig.~\ref{fig:pulse_temperature}. These phonon arrival time histograms do not directly map to the observed pulse shape, as they must be convolved with the QP lifetime and the resonator ring-down time. Additionally, it is clear from the time constants extracted from the pulse shapes, which range into the tens of milliseconds, that there are effects unmodeled in the simulation that extend the phonon arrival times (potentially, phonon recycling). 

\section{Improving Resolution of Current Architecture KIPM Detectors} \label{sec:near_term_improvements}

The resolution on energy deposited on the substrate, as measured by $N_r$ resonators (assuming the same resolution on energy absorbed for each resonator), is given by
\begin{align}
    \sigma_{E_\mathrm{dep}} = \frac{\sqrt{N_r}}{\eta(N_r)} \sigma_{E_\mathrm{abs}}~,
\end{align}
where $\eta(N_r)$ is taken to be the collection efficiency across all $N_r$ resonators on a device and $\sigma_{E_\mathrm{abs}}$ is the resolution on absorbed energy of each KID individually. Upon first inspection, going to multi-resonator devices worsens the device resolution as $\sqrt{N_r}$, but that is only true if $\eta$ is unchanging with $N_r$. By increasing resonator coverage, provided $\eta(N_r)$ grows faster than $\sqrt{N_r}$, we continue to improve $\sigma_{E_\mathrm{dep}}$ until becoming limited by loss, which the SuperCMDS collaboration has shown to occur at single-percent surface coverage~\cite{supercdms_fdr}. 

With aluminum resonators of our current architecture, we are thus motivated to return to multi-resonator devices: an improved version of Fig.~\ref{fig:device-masks} (A). First, we note our current design is limited by TLS noise, setting the achieved $\sigma_{E_\mathrm{abs}}=2.1$ eV. Were this noise not present, the equivalent HEMT-amplifier-limited energy resolution is improved by a factor four, both due to the decreased noise power and the improvement in the performance of the OF because the noise is white. With no concrete path to reduce the TLS noise, we may instead evade it by switching to dissipation readout~\cite{Gao2008c,Gao2007}, which suffers from $\approx$4$\times$ lower responsivity to $n_\mathrm{qp}$ than frequency readout. We can recover this factor by deploying a kinetic-inductance traveling-wave parametric amplifier (KI-TWPA) for first-stage amplification (with noise temperature $T_\mathrm{N} = 0.2-0.3$~K, $\approx$20$\times$ lower than a HEMT amplifier), which improves the resolution by a factor of 4--5, as demonstrated in~\cite{Ramanathan2024a}. We project the best-achievable sensor resolution of our current design to be $\sigma_{E_\mathrm{abs}}=500$ meV. By comparison, for an inductor volume of 12,000 $\mu$m$^3$ (\textit{e.g.}, Mask C), the energy resolution limit set purely by generation-recombination (GR) noise is $\sigma_{E_\mathrm{abs}}^\mathrm{GR-limit}=195$ meV. Using a low-$T_c$ material for the inductor (\S\ref{sec:lowTc}) is projected to reduce both by another factor four, yielding an ultimate projection of $\sigma_{E_\mathrm{abs}}=125$ meV for one resonator.

What remains is improving $\eta$, limited by loss to sub-gap phonons to $\eta_\mathrm{max}=0.46$~\cite{Wen2025}. We frame the remainder of this discussion using a predictive model for phonon absorption of specific detector elements based on the volume of superconductor present~\cite{Wen2025}. This scaling law codifies two important facts about $\eta$. First, when there is enough metal, losses to device mounts can be reduced by adding more metal (active or inactive). Second, for low mounting loss to result in useful $\eta$, the bulk of that metal must be active. First, we will implement wire-bond suspension~\cite{AnthonyPetersen2022} for device mounting, partially mitigating this loss. Next, we replace the Al interdigitated capacitor (IDC) with Nb, which is less efficient than Al at collecting phonons, and halve the thickness of the feedline to offset the volume increase from extending the feedline length to accommodate more resonators.  Quantitatively, the scaling law tells us that a 33-KID design (2\% surface coverage) using Nb IDCs will obtain $\eta = 27\%$ with negligible mounting losses, resulting in a detector with $\sigma_{E_\mathrm{dep}}\approx2.7$ eV, a factor of ten better than the state of the art~\cite{Delicato2023} with a target mass $\approx1$ g.

We further envision a larger format detector of 50 resonators on a 3.5-cm-diameter, 12-mm-thick Si substrate (27 g) using the same improvements. In this case, the surface coverage fraction of our detector remains the same as in the above case, thus so does $\eta$, yielding a detector resolution of $\sigma_{E_\mathrm{dep}} \approx 3.3~\mathrm{eV}$. Fashioning this into a phonon-only iZIP-style detector~\cite{SuperCDMSCollaboration2022} enables discrimination of nuclear and electronic recoils for DM masses down to $m_\chi = 1.5~\mathrm{GeV}$. Using a Ge substrate, which degrades $\eta$ by $1.5\times$ (yielding $\sigma_{E_\mathrm{dep}}\approx5$ eV), this detector would reach the neutrino fog in 3.3 kg $\times$ yr~\cite{SuperCDMSCollaboration2022}, optimistically assuming the detector is not limited by the low-energy excess~\cite{Baxter2025}.
%provide a $4\sigma$ detection of a 20 eV nuclear recoil.
% 

\section{Low-$T_c$ Resonators} \label{sec:lowTc}
\noindent As KIDs are sensitive to changes in the QP density $\delta n_\mathrm{qp}$, one can increase the device responsivity for the same absorbed energy by lowering the superconducting gap $\Delta$, or equivalently $T_c$. The observable quantity, the resonant frequency shift, is given by
\begin{align}
    \frac{\delta f_r}{f_r} = - \frac{1}{2} \alpha\ \kappa_2\ \delta n_\mathrm{qp} =  - \frac{1}{2} \alpha\ \kappa_2(\Delta) \frac{E_\mathrm{abs} }{V\ \Delta}~, 
\end{align}
where $\alpha$ is the kinetic inductance fraction and $\kappa_2$ is the imaginary part of the fractional change in complex conductivity per unit change in quasiparticle density. Thus lowering $\Delta$ leads to a higher energy responsivity, with all other quantities fixed, due to the lower-energy quanta. This motivates increasing the device energy resolution by designing resonators from low-$T_c$ materials, including Hf, Ir, and AlMn. Furthermore, one can use interfaces of materials with differing $T_c$ to exploit the phenomenon of QP trapping~\cite{Saab2000,Cabrera2000}. 

The LBNL team has recently fabricated resonators using hafnium, with a $T_c < 250$ mK~\cite{Li2025} ($2\Delta < 76~\mu$eV) and $Q_i>10^5$. Phonon pulses were observed in each resonator of a two-channel device, enabling position reconstruction based on both the partitioning of phonon energy between the two and the difference in pulse arrival times. Evaluation of the device's energy resolution and the performance of Hf resonators with Al phonon absorbers is left to a future publication.

Similarly, consortium members in the Materials Sciences Division at Argonne National Laboratory recently produced single-layer iridium resonators using the Mask C KIPM detector design. Iridium is an attractive material for KIPM detectors as it has no native oxides, enabling more efficient quasiparticle trapping (and potentially lower TLS noise). A 30-nm-thick witness sample of Ir from this fabrication run showed a $T_c$ of 285 mK and a sheet resistance of 4.95 $\Omega/\square$. The SLAC team has fabricated AlMn devices on a Si substrate, demonstrating $T_c=0.5$ K with $Q_i >10^{6}$. We plan to make both monolithic and trapping-enabled KIPM detectors using AlMn.

\section{Phonon-Absorber-Assisted Kinetic Inductance Phonon-Mediated Detectors} \label{sec:paakipm}
While the understanding we have obtained of our current KIPM architecture provides a clear near-term path to eV resolution, the ultimate limit posed by GR noise is
\begin{align} \label{eq:qp_E_res}
    \sigma_{E_\mathrm{dep}} = \frac{V \Delta}{\eta} \sigma_{n_\mathrm{qp}} = 
    \frac{\Delta}{\eta} \sqrt{2 \pi n_\mathrm{qp}^* V}
\end{align}
In the current architecture, reducing GR noise further requires reducing $V$, which inevitably also reduces $\eta$, as we have seen. We are thus developing a different architecture, the Phonon-Absorber-Assisted (PAA) KIPM detector, that uses QP trapping to decouple $V$ from $\eta$, taking inspiration from the success of the quasiparticle-trapping-assisted electrothermal-feedback TES (QET)~\cite{Irwin1995a,Nam1996,Chang2025}. The challenge has been to identify a design that is consistent with RF design requirements, which we have now found.

Like the current KIPM design, the PAA-KIPM detector uses many individual sensors, which we term PAA-KIDs.  The inductor of a PAA-KID consists of a series array of many large area (100$\times$100 $\mu$m$^2$), thick (600 nm) Al phonon absorbers to obtain $\eta$ and an equal number of narrow (1-2 $\mu$m), short (2-10 $\mu$m), and thin (30-100 nm) segments of lower-$T_c$ material to reduce $V$ substantially.  The low-$T_c$ segments have much higher kinetic inductance per square than the Al because they are thin films, and they also dominate the number of squares because they are narrow. A parallel-plate capacitor (PPC) with low-noise hydrogenated amorphous silicon (a-Si:H) dielectric, developed by members of our consortium~\cite{Defrance2024, Golwala2024}, completes the resonator, providing high capacitance per unit area and thus minimizing the inactive capacitor area. 

A key consideration in the PAA-KID design is the choice of low-$T_c$ material. A lower $\Delta$ yields a smaller coefficient in Eq~\ref{eq:qp_E_res} and improves trapping efficiency but could also result in a higher $n_\mathrm{qp}^*$ or a lower limit on readout power, which can degrade amplifier and/or TLS noise. The normal-state resistance per square of the low-$T_c$ material is also important, as higher values yield higher kinetic inductances and thus higher responsivity, but also a lower bifurcation power (onset of KI nonlinearity). We analyze the design in detail in~\cite{XiongLTD2025}, yielding an expected PAA-KID resolution of $\mathscr{O}$(1 meV) and, with 4\% area coverage providing $\eta$ = 35\%, a PAA-KIPM detector resolution on deposited energy of $\mathscr{O}$(10 meV). A gram-month exposure with such a detector will probe a DM-electron cross section of $\mathscr{O}(10^{-24}$ cm$^2$) at $m_\chi=$ 50 keV, a mass range that cannot be probed with non-phonon-mediated detectors. The PAA-KID architecture would enable construction of a phonon-only iZIP-like detector to probe into the neutrino fog at 0.3--5 GeV. Again, these projections do not consider the low-energy excess phonon-only background.

\section{Consortium Facilities \& Capabilities} \label{sec:capabilities}

Our consortium has access to six nanofabrication facilities through their member institutions or close collaborators: the MicroDevices Laboratory at the Jet Propulsion Laboratory, the Stanford Nanofabrication Facility, the Argonne Materials Sciences Division, NuFAB at Northwestern University, the Pritzker Nanofabrication Facility at the University of Chicago, and the LBNL Molecular Foundry. %Between these facilities, w
We have the capability of depositing standard superconductor materials (Al, Nb, etc.), as well as the three low-$T_c$ materials discussed above. While the current Hf resonators under test were fabricated from wafers purchased from StarCryo, LBNL is developing Hf deposition and fabrication processes. The Stanford Nanofabrication Facility has the capability of working with large-format substrates.

Within the consortium, we have at our disposal 11 $^3$He/$^4$He dilution refrigerators (DRs) at 5 institutions in the US. 

\subsubsection{NEXUS}

The Northwestern EXperimental Underground Site is a low-background cryogenic test stand located in the MINOS cavern at Fermilab with a 225 m.w.e. rock overburden~\cite{Michael2008}, providing a $>$99.5\% reduction in the cosmogenic muon rate. The facility features a movable, 4''-thick lead shield that encloses a CryoConcept HEXA-DRY DR. A lead plug interior to the DR, above the detector payload volume, completes the $4\pi$ shielding and provides a 100$\times$ reduction of the ambient radiogenic gamma background. The facility also houses an Adelphi DD108 deuterium-deuterium fusion generator, a source of 2.45 MeV monochromatic neutrons, which, in conjunction with an in-development, segmented backing array of scintillator-based detectors, enables fixed-angle scattering experiments for outgoing neutron angles $<10^\circ$. Such experiments can perform nuclear-recoil calibrations down to 50 eV$_\mathrm{nr}$ in Si (for $1^\circ$ angular resolution).

\subsubsection{QUIET \& LOUD} Located in the same experimental cavern at Fermilab as NEXUS, the Quantum Underground Instrumentation Experimental Testbed is an Oxford Proteox DR. The experimental payload region is encased in a double-walled A4K magnetic shield. Near-term planned upgrades include the addition of a magnet capable of applying fields up to 8 T, as well as an ionizing radiation shield surrounding the experimental payload. The sister site to QUIET, LOUD is an identical facility located on the surface at Fermilab, intended for rapid throughput testing of superconducting devices. 

\subsubsection{SLAC Millikelvin Facility} The SLAC group operates an Oxford Proteox on the surface with an identical magnetic shield as in QUIET. This facility features a cryogenic scannable optical calibration source based on a MEMS mirror~\cite{Stifter2024,Tabassum2025}. The facility also has 3 BlueFors LD400, 1 BlueFors XLD400, a $^3$He sorption fridge, and an additional Oxford Proteox, all available for KIPM testing.

\subsubsection{Other facilities} At the Washington University in St. Louis, we have at our disposal a Leiden DR as well as a 4K testbed for rapid testing of devices with high $T_c$ structures. The Caltech group operates a Oxford KelvinOx 25 wet DR which, due to the lack of a pulse tube, provides a low-vibration testing environment important for devices using wire-bond suspension~\cite{AnthonyPetersen2022,AnthonyPetersen2024,Chang2025}. The team at LBNL/UCB has two Bluefors LD400 DRs with optical photon calibration fibers and magnetic shields available for KIPM detector deployment.

\section{Conclusion} \label{sec:conclusion}
The KIPM Detector Consortium is advancing phonon-mediated detectors for low-energy rare event searches, primarily DM direct detection. The experiments and simulations discussed in this work used single-resonator devices and a focused light source, and exhibit position-dependence of the phonon collection efficiency. While this implies $\sigma_{E_\mathrm{dep}}$ quoted for single-resonator devices is not applicable to diffusive sources (including DM), ultimately, the Consortium is moving to multiple-resonator devices with high surface coverage (\S\ref{sec:near_term_improvements}). The position dependence in such a detector will be much more modest due to the larger density of sensors. Furthermore, this will allow for position reconstruction~\cite{Moore2012a}, and correction for the position dependence of $\eta$.

Near-term improvements to the resonator and device design optimizing for phonon collection efficiency are projected to enable deployment of multi-resonator detectors with $\sigma_{E_\mathrm{dep}}=2.7(3.3)$ eV and a target mass of 1(27) grams. To reach meV-scale resolutions, we are developing a new resonator architecture that incorporate quasiparticle trapping into low-$T_c$ materials: the PAA-KIPM detector, with a projected single-resonator energy resolution of $\sigma_{E_\mathrm{abs}}=\mathscr{O}(1$ meV). 
These advanced designs, which must be validated experimentally, coupled with an evolving understanding of pulse shape make KIPM detectors an exciting option for sub-GeV DM searches and other applications.

\section{Acknowledgments}
\noindent The authors acknowledge Professors Daniel Baxter and Enectali Figueroa-Feliciano for their support of NEXUS and access to NUFab. This document was prepared by the KIPM Detector Consortium using the resources of the Fermi National Accelerator Laboratory (Fermilab), a U.S. Department of Energy, Office of Science, Office of High Energy Physics HEP User Facility. Fermilab is managed by FermiForward Discovery Group, LLC, acting under Contract No. 89243024CSC000002. Partial funding was furnished by the FNAL Laboratory Directed Research \& Development program award number LDRD2020-040, NASA NST GRO80NSSC20K1223, and the Department of Energy DE-SC0011925F. This work was supported by the U.S. Government under ARO grant W911NF-22-1-0257 and the U.S. Department of Energy, Office of Science, National Quantum Information Science Research Centers, Quantum Science Center and Q-NEXT and the U.S. Department of Energy, Office of Science, High-Energy Physics Program Office. 

\subsection*{Author Affiliations}
D.J. Temples is with the Emerging Technologies Directorate, Fermi National Accelerator Laboratory, Batavia, IL, USA.

W. Ho and K. Ramanathan are with the Department of Physics, Washington University in St. Louis, St. Louis, MO, USA.

T. Aralis, S.Q. Dang, N. Kurinsky, Z. Smith, and O. Wen, are with the Kavli Institute for Particle Astrophysics and Cosmology, SLAC National Laboratory, Menlo Park, CA, USA. S. Q. Dang and Z. Smith are also with the Department of Physics, Stanford University, Stanford, CA, USA.

Y.-Y. Chang, M. Garcia-Sciveres, K. Li, X. Li, A. Suzuki, and M. Williams are with the Division of Physics, Lawrence Berkeley National Laboratory, Berkeley, CA, USA.

S. Golwala, C. Cap, S. Ray, B. Sandoval, and J.R. Xiong are with the Division of Physics, Mathematics, and Astronomy, California Institute of Technology, Pasadena, CA, USA.

G. Wang, M. Lisovenko, and V. Yefremenko are with the Division of Physical Sciences \& Engineering, Argonne National Laboratory, Lemont, IL, USA.

C. Chang is with the Kavli Institute for Cosmological Physics, University of Chicago, Chicago, IL, USA.

E. Panner is with the Department of Physics and Astronomy, Tufts University, Medford, MA, USA.

\bibliographystyle{IEEEtran}
\bibliography{main.bib}

\end{document}